\documentclass[11pt,preprint]{aastex}

\shorttitle{Variable absorption in NGC 1365}
\shortauthors{G. Risaliti et al.}

\begin{document}

\title{Variable partial covering and a relativistic iron line in NGC~1365}

\author{G. Risaliti\altaffilmark{1,2},
G. Miniutti\altaffilmark{3,4}
M. Elvis\altaffilmark{1}, 
G. Fabbiano\altaffilmark{1}, 
M. Salvati\altaffilmark{2},
A. Baldi\altaffilmark{1},
V. Braito\altaffilmark{5,6},
S. Bianchi\altaffilmark{7},
G. Matt\altaffilmark{7},
J. Reeves\altaffilmark{8},
R. Soria\altaffilmark{9},
A. Zezas\altaffilmark{1}
}
\email{grisaliti@cfa.harvard.edu}

\altaffiltext{1}{Harvard-Smithsonian Center for Astrophysics, 60 Garden St. 
Cambridge, MA 02138 USA}
\altaffiltext{2}{INAF - Osservatorio di Arcetri, Largo E. Fermi 5,
I-50125, Firenze, Italy}
\altaffiltext{3}{
Laboratoire Astroparticule et Cosmologie (APC), UMR 7164, 10 Rue A.
Domon et L. Duquet, 75205 Paris}
\altaffiltext{4}{LAEX, Centro de Astrobiologia (CSIC-INTA); LAEFF, P.O. Box 78, E-28691, Villanueva de la Ca\~nada, Madrid, Spain}
\altaffiltext{5}{Department of Physics and Astronomy, University of Leicester, University Road, Leicester, LE1 7RH, UK}
\altaffiltext{6}{Department of Physics and Astronomy, Johns Hopkins University, Baltimore, MD 21218}
\altaffiltext{7}{Dipartimento di Fisica, Universit\`a degli Studi ``Roma Tre'',
Via della Vasca Navale 84, I-00146 Roma, Italy}
\altaffiltext{8}{Astrophysics Group, School of Physical and Geographical Science, Keele University, Keele, Staffordshire ST5 5BG, UK}
\altaffiltext{9}{MSSL, University College London, Holmbury St. Mary, Dorking, Surrey, RH5 6NT, UK}

\begin{abstract}

We present a complete analysis of the hard X-ray (2-10~keV)
properties of 
the Seyfert galaxy NGC 1365, based on a 60 ks {\em XMM-Newton} observation
performed in January 2004.
The two main results are: 1) We detect 
an obscuring cloud with $N_H\sim3.5\times10^{23}$~cm$^{-2}$
crossing the line of sight 
in $\sim25$~ks.
This implies a dimension of the X-ray source not
larger than a few 10$^{13}$~cm and a distance of the obscuring cloud
of the order of 10$^{16}$~cm. Adopting the black hole mass $M_{BH}$ estimated
from the $M_{BH}$--velocity dispersion relation,
the source size is $D_S<20~R_G$ and the distance and density of the obscuring
clouds are $R\sim3000-10000~R_G$ and $n\sim10^{10}$~cm$^{-3}$, i.e. 
typical values for broad line region clouds.  2) An iron emission line
with a relativistic profile is detected with high statistical significance.
A time integrated fit of the line+continuum reflection components
suggests a high iron abundance ($\sim3$ times solar) and an origin of
these components in the inner part (within $\sim10~R_G$) 
of the accretion disk, in agreement with the small source size inferred from
the analysis of the absorption variability. 
\end{abstract}


\section{Introduction}

According to the Unified Model of Active Galactic Nuclei (AGNs, see
review by Urry \& Padovani, 1995), an axisymmetric
absorber/reflector is present around the central black hole of AGNs,
with a size between that of the Broad Emission Line Region (BELR,
$\sim 1000$ Schwarzschild radii) and that of the narrow line region
(tens to hundreds parsecs).  The easiest geometrical and physical
configuration of such absorber is that of a homogeneous torus on a
parsec scale (Krolik \& Begelman 1988).
However, this view has been recently challenged by several pieces of
observational evidence:
\\  
(1) Dramatic X-ray absorbing column density changes (factors of $>$10)
over a few years have been seen in several type~2 (narrow permitted
lines) Seyferts (Risaliti, Elvis \& Nicastro 2002), 
ruling out a homogeneous absorber.  \\
(2) Rapid column density variability, on time scales of hours,
requires an X-ray absorber no larger than the BELR. Such changes have
been detected in the brightest absorbed Seyfert galaxy, NGC~4151
($\sim10-30$~hours, Puccetti et al. 2004) and in the Seyfert~2 galaxy NGC~4388
(4~hours, Elvis et al. 2004). Assuming that the absorber is made by
material moving with Keplerian velocity around the central source,
these observations imply that its distance is of the order of
the broad line clouds, i.e. a small fraction of a parsec.  In order to
have the absorber at a parsec from the center, 
the linear size of the clouds should be of the order of
10$^{11}$~cm, much smaller than the minimum possible size of the 
X-ray source (at least $\sim10$ gravitational radii, i.e. several 10$^{12}$~cm
for a $10^7~M_\odot$ black hole).
\\ 
(3) An indication of strong inhomogeneities in the absorber/reflector
comes from the analysis of the reflection components in the X-ray spectra of Compton-thin
Seyfert Galaxies, which are systematically stronger than expected from
reflection by gas with the same column density as measured in
absorption. This has been shown convincingly for a few sources in
which a detailed measurement of the reflection component has been
possible (NGC~1365, Risaliti et al. 2000; NGC~2992, Gilli et
al. 2002; NGC~6300, Guainazzi et al. 2003) and in a statistical sense
in a sample of $\sim 20$ bright Seyfert 2s (Risaliti 2002).
This implies the presence of Compton-thick reflecting material
out of the line of sight, and covering a significant fraction of
the solid angle as seen from the X-ray source (Ghisellini, Haardt \& Matt
1994, Magdziarz \& Zdziarski 1995).

Here we report on a 60-ks {\em XMM-Newton} observation of the 
obscured AGN in NGC~1365, which put new stronger constraints
on the size and structure of its circumnuclear absorber.

The same observation also revealed a prominent broad iron emission line
with a relativistic profile.
Asymmetric profiles of iron-K$\alpha$ emission lines are one of the possible
probes of the extreme general relativistic effects around black holes.
This issue has been thoroughly investigated in the past 15-20 years
both from a theoretical and an observational point of view.
The iron emission from an accretion disk around a black hole has been
modeled taking into account the broadening and Doppler boosting due to the disk rotation,
and the gravitational redshift due to the black hole field (Fabian et al.~1990).
Several models predict the line profile from a maximally rotating black hole
(e.g. Laor~1991, Dovciak et al.~2004), considering different emissivity
profiles, emitting regions, disk ionization states and viewing angles
(for a review, see Fabian \& Miniutti~2005). 
Observationally, the asymmetric profile attributed to the relativistic
line has been first observed by ASCA in MCG-6-30-15 (Tanaka et al.~1995),
and subsequently in several AGNs and Galactic black holes 
with ASCA, {\em BeppoSAX}, {\em XMM-Newton}, and
{\em Suzaku} (e.g. Miller~2007, Nandra et al.~2007)

The main open issue in this field is due to the large width of the relativistically
broadened line, extending from the peak energy around 6-7~keV down to
3-5~keV in the most extreme cases, depending on the disk parameters. Such a broad feature can be
effectively mimicked by continuum curvatures, due to reflection components,
partial covering by circumnuclear gas, or to a poor determination of the
intrinsic continuum (usually modeled with a simple power~law).
Recently, Miller, Turner \& Reeves~(2008) showed that it is formally possible to 
reproduce the X-ray spectrum of MCG-6-30-15, with its broad
emission feature (which is so far the highest
quality relativistic line candidate) through a model which includes a
warm absorber with a variable partial covering of the central source.

Here we show that the identification of the broad feature 
in the {\em XMM-Newton} spectrum of NGC~1365 as a relativistically
broadened iron-K line is rather solid, thanks to a combination
of favourable aspects: a high iron metallicity, which implies a
large equivalent width of the line, and a sharp low-energy tail, which is impossible to
reproduce with partial covering scenarios. 

%
%
%
%
\section{The X-ray observational history of NGC 1365}

NGC~1365 ($z$=0.0055) is a particularly striking example of absorption variability. 
It was observed by ASCA in
1995 in a reflection-dominated state (N$_H>$10$^{24}$~cm$^{-2}$, Iyomoto et
al. 1997) and, three years later, in a Compton-thin state
($N_H$$\sim$4$\times$10$^{23}$~cm$^{-2}$) by BeppoSAX (Risaliti et
al. 2000).  The ratio between the 2-10 keV flux of the reflected component
and that of the intrinsic spectrum was higher than 5\%. Such
a high reflection efficiency can be achieved only if a thick
(N$_H>3\times 10^{24}$~cm$^{-2}$) reflector covers a large fraction of
the solid angle around the central source (Ghisellini
et al.~1994, Magdziarz \& Zdziarski 1995). 

The switch from reflection-dominated to Compton-thin states could in principle be due either to
extreme variations of the intrinsic luminosity (Matt, Guainazzi \& Maiolino 2003)
or to Compton-thick clouds moving across the line of sight.

%
In order to resolve the uncertainty between these two physical explanations,
and to study the complex structure of the circumnuclear material,
NGC~1365 was observed many times with {\em XMM-Newton} and {\em Chandra} in the
past few years, revealing a series of unique properties:


%
(1) The X-ray obscuration varies dramatically. We first found two
Compton-thick to Compton-thin transitions in a series of three {\em Chandra} and
{\em XMM-Newton} snapshot observations performed in a total period of six weeks
(Risaliti et al.~2005a, hereafter R05A).  We then performed a {\em Chandra} monitoring
campaign consisting of six short (10~ks) observations in ten days, which revealed
changes from
reflection-dominated to transmission-dominated
spectra on a time scale
shorter than two days (Risaliti et al.~2007).
This fast variability completely rules out the 
intrinsic variability scenario, leaving occultation by a circumnuclear cloud as the
only plausible cause of the observed variations.\\
(2) A series of four {\em XMM-Newton} observations, during which the source was
in a Compton-thin state, revealed a complex of absorption lines, detected 
with high statistical significance, at energies between 6.7 keV and 8
keV (Risaliti et al.~2005b, hereafter R05B). 
These lines are due to FeXXV and FeXXVI $K\alpha$ and $K\beta$ resonant absorption
by a highly ionized gas with column density of a few 10$^{23}$~cm$^{-2}$, one of
the most extreme warm absorbers (both for the high $N_H$ and the high ionization state)
detected so far among AGNs.\\
(3) The soft thermal component which dominates below 2~keV is constant within 5\% in all
observations.  The high spatial resolution of {\em Chandra} allowed us to
resolve the emitting region, which extends over $\sim 1$~kpc from the
center, while the hard component originates in a region with diameter $<100$~pc
(10~arcsec), as commonly observed in Seyfert Galaxies (Bianchi et al.~2006, Guainazzi \& Bianchi~2007).
The analysis of the {\em Chandra}  observations of spatially diffused component
is presented in a companion paper (Wang et al.~2009).
\\
%


\section{Data Reduction}

The observation was performed on January 18-19 2004, with
the EPIC PN and MOS instruments (Str{\"u}der et al.~2001, Turner et al.~2001) 
onboard the {\em XMM-Newton} observatory.
Simultaneous RGS spectra were also obtained. These data  have been presented by Guainazzi \& Bianchi (2007) and
are not discussed in
the present work.
The data were reduced using the SAS package following 
the standard procedure suggested in the {\em XMM-Newton} manuals and
threads\footnote{http://xmm.esac.esa.int/sas/8.0.0/}. The data were first checked for high background flares.
No strong flare was present, so the whole exposure 
has been used in the analysis.
The MOS event files were merged before the extraction of the scientific data.
The PN data were treated separately.
 
Spectra and light curves were extracted from a 30 arcsec radius circular region around
the nuclear source. 
The EPIC image is complex, showing an extended galactic emission
and several individual galactic sources, but the nuclear source within 30 arcsec
outshines the rest of the galaxy by a factor $>30$.
The background was extracted from nearby regions (but outside the galaxy) in the same
fields. The response files were created using the SAS
package.

The same procedure has been repeated for all the subsequent extractions
where time or energy filters have been applied (as we will show later, 
we extracted light curves in 
specific energy ranges, and spectra in many different time intervals).
The relevant data about the spectra obtained from the whole observation,
and from the different time intervals discussed in the next Sections,
are shown in Table~1.

\begin{table}
\caption{NGC 1365 - General parameters}
\centerline{\begin{tabular}{lccc}
\hline
OBS   &Counts& T$^a$ & F(2-10)$^c$ \\
\hline
TOTAL &103907& 60 &  1.37 \\
INT 1 &41784& 22 &1.39 \\
INT 2 &44362& 26 & 1.20 \\
INT 3 &17761& 12 &  1.42 \\
\hline
\end{tabular}}
\footnotesize{
$^a$: Observing time, in ks. $^b$:  observed 2-10~keV flux, in units of $10^{-11}$~erg~s$^{-1}$~cm$^{-2}$. }
\end{table}

\section{Data Analysis: full observation}

The analysis of the spectra has been partially presented in a previous paper
(R05B) where we focused on one single spectral feature:
the presence of four iron absorption lines in the 6.7-8.3~keV spectral interval.
Since these are narrow features, it was possible to investigate their properties
without discussing the continuum fit in detail: any analytical fit correctly representing
the continuum level at the lines' energy was acceptable.

Here we are instead interested in the physical properties of the 2-10~keV continuum and in broad 
features. 

A second observation of similar length was performed six months later,
and is also presented in R05B. We performed a complete analysis of
these data, analogous to the one presented here.
No significant spectral variability was detected
during this observation. Moreover, due to the 
high measured column density ($N_H\sim4\times10^{23}$~cm$^{-2}$)
most of the iron line broad emission is obscured.
For these reasons we do not discuss all the details of
the data analysis. A summary of the relevant results 
is presented in Section~5.5.

The spectral analysis has been performed using the XSPEC~12.4 analysis package (Arnaud et al.~1996).
In all cases we used both the PN and MOS spectra. We normalized our fluxes to the
MOS values. The PN normalizations are always consistent with those of MOS within $<$5\%.
All the parameters and errors quoted in this paper refer to simultaneous fits to 
both PN and MOS data. However, many of the figures contain PN data only, for the sake of clarity.

The model adopted in R05B consists of the following components:\\
1. a continuum power law absorbed by a column density $N_H\sim2\times10^{23}$~cm$^{-2}$\\
2. a set of four absorption lines at energies between 6.7 and 8.3~keV, due to iron XXV and XXVI
$K\alpha$ and $K\beta$ transitions.\\
3. a cold reflection component modeled through the PEXRAV model (Magzdiarz \& Zdziarski~1995).\\
4. an unabsorbed power law, with  a normalization of a few percent of that of the main continuum component, and
the same photon index.\\
5. a broad FeK iron emission line, with a relativistic profile.\\
6. a narrow (unresolved) iron FeK emission line, at rest-frame energy E=6.39~keV.\\
7. an optically thin thermal component with kT$\sim0.8$~keV, responsible for the bulk of the soft ($E<2$~keV) emission.\\

In addition to the spectral components already listed, we also added 
a continuum from an ionized reflector (PEXRIV model
in XSPEC). This component is needed to model a possible second screen located close to the source,
(e.g. the accretion disk itself), 
and therefore likely
to be obscured when the source is in a Compton-thick state. Given the possible low distance
from the central black hole of the ionized gas, we also allowed for relativistic blurring of this component.
Specifically, 
we used the convolution model {\em kdblur} in XSPEC, developed by A. Fabian and R. Johnstone,
which smoothes an arbitrary continuum according to the relativistic effects from an accretion disk around a
maximally rotating black hole. 
The parameters of this component (inner and outer radius, and emissivity profile) are 
not tied to the corresponding parameters in the broad emission line component.
This in principle allows for different emitting regions for the line and continuum components. 
Since in none of the fits presented in the following Sections these parameters are significantly
constrained, we do not discuss them in the intermediate steps of our analysis, and we link
them to the broad emission line parameters in the final fit presented in Section 5.4.

The continuum reflection component (3) was initially
 frozen at the best value obtained from a previous {\em XMM-Newton} observation (OBS~2 in R05A)
which caught
the source in a Compton-thick, reflection-dominated state. This choice is based on the assumption
that the reflection component remains constant on long time scales. This is the case if the
distance of the cold reflector is large compared with the speed of light times the typical
variability time scale. This hypothesis could not be verified with the data analyzed here. However, the fast
absorption variability found with the {\em Chandra} campaign suggests that the absorber/reflector
is located quite close to the central source (distance $<10^{16}$~cm, Risaliti et al.~2007).
We will discuss possible variations of this component in the next Sections. Here we are
only interested in defining a baseline model to use as an initial step to develop our time-resolved analysis.

A complete analysis of the four iron absorption lines is presented in R05B, and is not repeated 
in this paper. Here we only mention the results relevant for the present analysis:\\
- the EW of the absorption lines is among the highest ever detected in AGNs 
(EW(Fe XXV K$\alpha$)$\sim$180~eV).\\
- The lines are partially saturated. This implies that a global fit with ionized absorber models
(e.g. through the XSTAR code) are not easily implemented. Instead, we analysed the lines parameters,
measured through Gaussian fits, with a model developed by Bianchi et al.~(2005), 
which takes into account line saturation effects, the ionization state of the obscuring gas, 
and its turbulence velocity.\\
- The result of our analysis is that absorber has an ionization parameter U$>$1000,\footnote{Throughout this paper,
the ionization parameter is defined as U=$L_X$/($n R^2$) where $L_X$ is the ionizing
X-ray luminosity and $n_e$ is the number density of particles in the gas. U is therefore expressed in units
of erg$\times$cm/s.} and
a column density $N_H\sim$a few~$10^{23}$~cm$^{-2}$.

In order to include possible effects of continuum absorption due to this gas, we used XSTAR
to estimate the global effect of this warm absorber on the continuum emission. 
We find that, due to the high ionization state, no other significant alteration of the incident
continuum (except for the iron absorption lines) is expected. Therefore, we always include the set
of four absorption 
lines in all the fits presented here, with no further absorption component due to this gas.

The model consisting of the components summarized above provided a marginally 
acceptable fit ($\chi^2=1905/1752$ degrees of freedom, Fig.~1). 


One of the most interesting features of this spectrum is the broad iron line (Fig.~2). 
This
component is well fit with a profile produced by a maximally rotating black hole (LAOR model
in XSPEC, Laor et al. 1991), plus a narrow (unresolved) component. Interestingly, the
best fit flux of the narrow component is consistent with the flux measured in the
purely reflection-dominated spectrum presented in R05A, suggesting a constancy of this component
on long (several years) time scales.

We note that several residuals
suggesting a non-optimal continuum fit are apparent in Figs.~1 and~2. This also implies
that the interpretation of the broad feature in Fig.~2 as a relativistic iron line has
to be considered as tentative.

In principle, we could try to obtain better results adopting a more complex
model, allowing for more absorption/reflection components. However, we
will show in the following that this is not the right approach in this case,
since most of the remaining deviations from a simple power law are due to
absorption variability.
 
\begin{figure}
\epsscale{0.6}
\includegraphics[width=11cm,angle=-90]{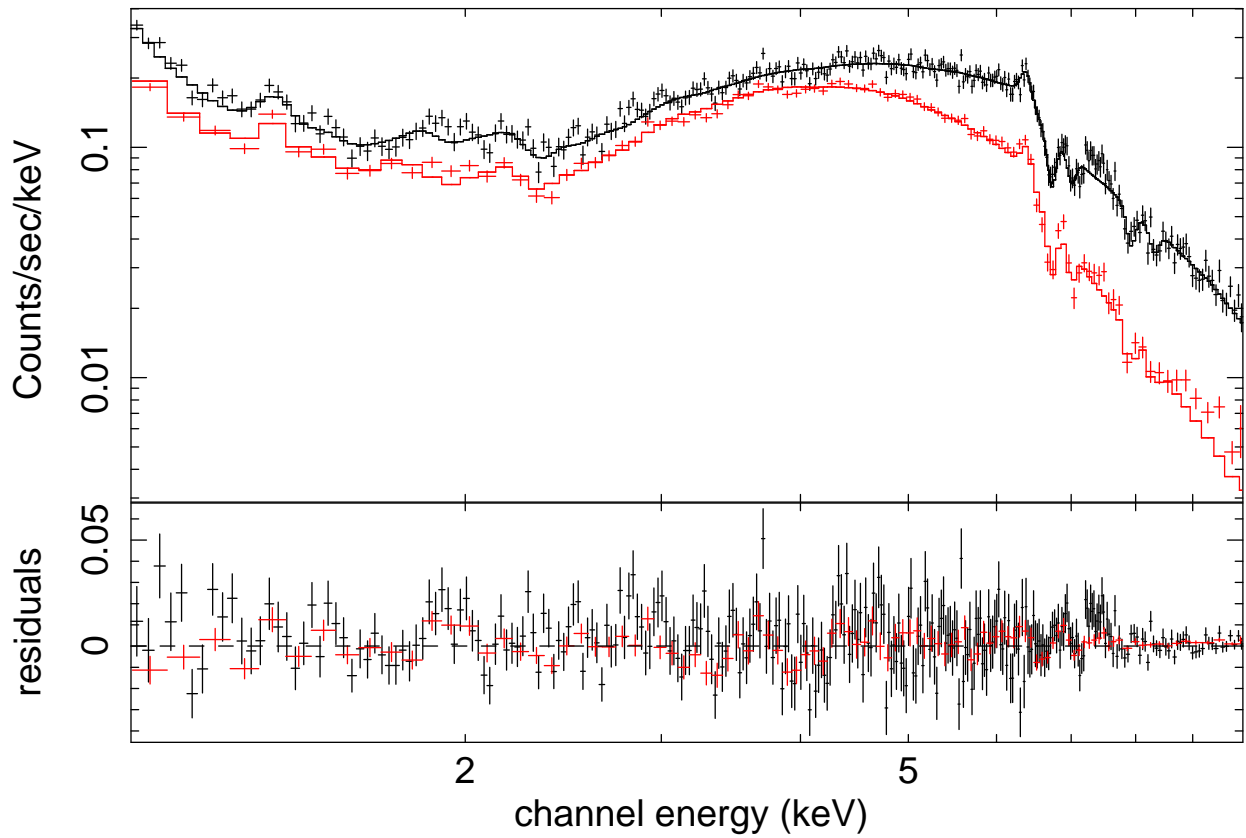}
\figcaption{\footnotesize{
Data, best fit model and residuals for spectrum obtained from the whole 60~ks
 {\em XMM-Newton} observation.
The baseline model described in the text, and first described in R05B, is adopted.
Both PN and MOS data are plotted. }}
\label{totfit}
\end{figure}
 
Here we start from the time integrated analysis to perform a more detailed, time-resolved study,
which takes into account possible spectral variations within the 60~ks total exposure.
In order to perform this study, we adopted a two step strategy: we first looked
for variations in the spectral shape during the observation, and sub-divided it 
in intervals with constant overall shape, then we
performed a complete
spectral analysis of each interval. 
\begin{figure}
\epsscale{0.9}
\plotone{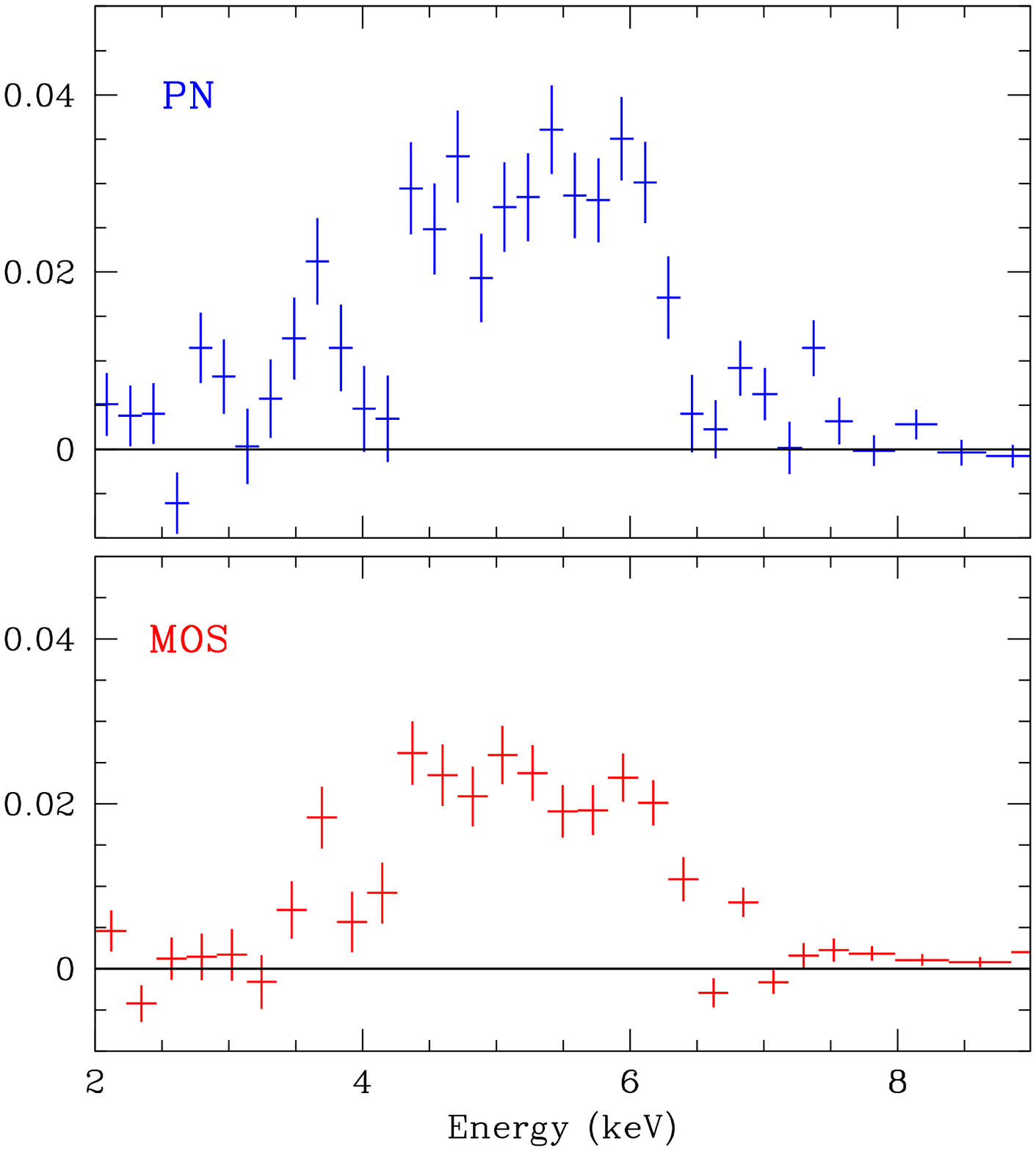} 
\figcaption{\footnotesize{
Residuals with respect to the best fit continuum model (without the iron emission features) described in the text.
Upper panel: PN instrument; lower panel: merged MOS. 
The fit was performed on the data without the 4-6.5~keV interval. These data were then re-noticed after the fit.
}}
\label{iron}
\end{figure}

We describe the details of this procedure in the next Sections.
\section{Time-resolved spectral analysis}

The simple analysis performed in R05B and summarized above shows that the primary emission
of NGC~1365 is absorbed at energies below 2-3~keV 
by a column density  $N_{H}\sim2\times10^{23}$~cm$^{-2}$. 
 Since we are interested in detecting absorption variations,
we analyzed the light curves of a hardness ratio $HR$ defined as the ratio between the 2-5~keV flux and
the 7-10~keV flux. The 2-5~keV spectral interval is where the photoelectric cutoff causes the strongest
changes in response to column density variations. The second interval is at energies well above the 
cut-off, and is not dependent on small $N_H$ variations. 
The $HR$ light curve is therefore optimized to detect $N_H$ variations, while it is insensitive
to intrinsic flux variations\footnote{This is not strictly true if a constant reflection component is
also present. However, the contribution of this component to the observed flux is at the level of a few percent,
therefore the effect on $HR$ is expected to be small.}. We note that a variation of the spectral shape of the
primary continuum would also imply a variation of $HR$. Therefore, changes in $HR$ do
not have  a strictly unique interpretation. However, this step does divide the total exposure into intervals with
constant spectral shape, which we then analyze separately. 
The $HR$ light curve is shown in Fig.~3.
A clear variability is detected, with $HR$ lower (i.e. harder spectrum) in the central $\sim$20~ks. 
The total 2-10~keV light curve, also plotted in Fig.~3, shows that on average the 
HR and total flux light curves are correlated. The flux light curve also shows variability on time
scales  shorter that those investigated with the $HR$ light curve. For such short variations,
a detailed spectral analysis is not possible. In the following we will only
investigate the variations well detected in the HR light curve.
We therefore extracted the spectra from three $\sim20$~ks intervals,
as indicated in Fig.~3.

We stress that this approach to spectral analysis is not an ad~hoc procedure to search 
for $N_H$ variations. Indeed, regardless of the physical reason for the observed variations,
 a time-resolved spectral
analysis is needed in order to correctly estimate  
the parameters of the continuum component. For example, adding states
with different intrinsic continuum slopes would introduce artificial curvatures in the spectrum, which
could be misrepresented with reflection components and/or broad emission features. Therefore, our method
can be considered a generalization of the standard time-averaged spectral analysis.

\begin{figure}
\epsscale{0.8}
\plotone{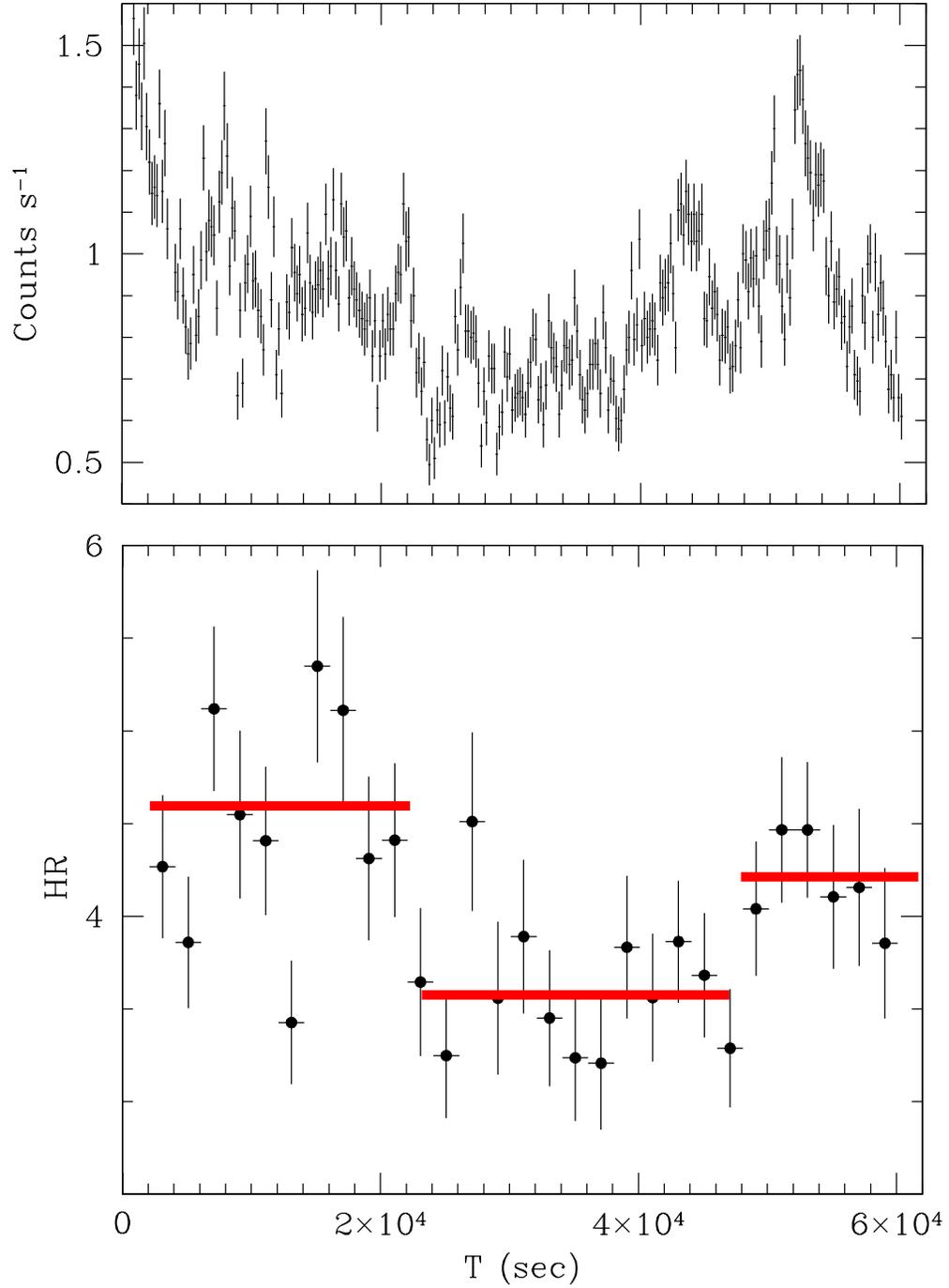} 
\figcaption{\footnotesize{Upper panel: 2-10~keV light curve of the AGN in NGC~1365 (bin size 100~s). Lower panel:
Hardness ratio $HR$ light curve $F(2-5)/F(7-10)$ (bin size 2,000~s).
The horizontal lines show the intervals chosen for the time-resolved spectral analysis.
}}
\label{lightcurves}
\end{figure}

\subsection{Independent spectral fits of each interval}

The spectra extracted from the three time intervals of Fig.~3 are plotted in Fig.~4.
From a visual inspection of these plots we immediately notice that the hardness
ratio changes in Fig~3 are due to a decrease of the $\sim2.5-6$~keV flux in the
second interval, while the three spectra are constant at higher energy.
This kind of variation is 
 suggestive of a column density variation, which affects only the energy range around
the cut~off, rather than a change in the spectral shape of the intrinsic emission.
In order to check this qualitative analysis, 
we performed a complete spectral analysis of each interval, using the best   
fit model of the complete observation as a baseline. At first we left all the parameters free, except for
the soft thermal component, 
which is constant within $<3$\% in all intervals, and is known to be emitted
by an extended component on a kiloparsec scale, resolved by {\em Chandra} observations (R05A).

\begin{figure}
\epsscale{0.8}
\plotone{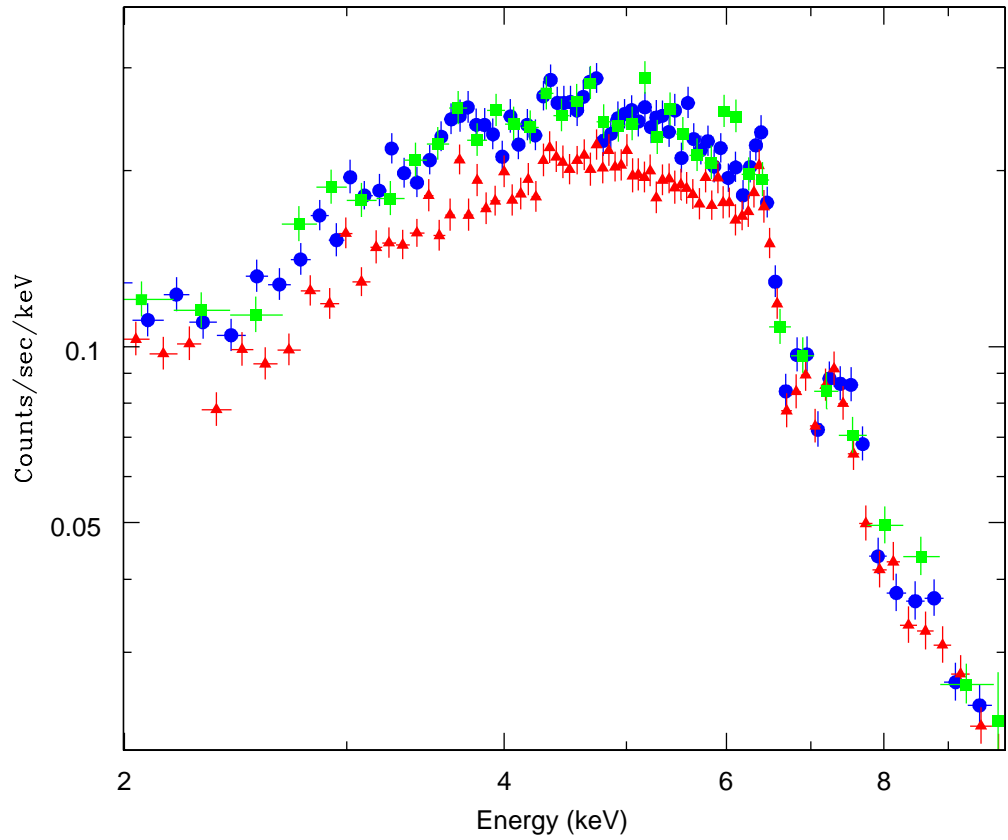} 
\figcaption{\footnotesize{
Spectra obtained from the three time intervals shown in Fig.~3. 
The spectra from the first and third interval (blue circles and 
green squares) are equal within the errors. The second spectrum 
(red triangles) is significantly different below $\sim$6~keV,
and compatible with the others at higher energies. Only PN data are plotted.
}}
\label{threespectra}
\end{figure}

The results are summarized in Table~2 and~3, 
together with the best fit parameters obtained in the time integrated fit.
As already mentioned at the beginning of the Section, 
in all fits the relativistically broadened iron line is included, and is strongly requested
by the spectral fit of both the total observation and each of the three intervals.
In all cases, a relativistic disk profile around a maximally rotating black hole is
preferred to both a simple Gaussian model and a relativistic profile for a non-rotating black hole.

\begin{table*}
\caption{NGC 1365 - Time-resolved Spectral Fit - Continuum components}
\centerline{\begin{tabular}{lccccccccc}
\hline
OBS   & $\Gamma$            & N$_H^a$         & A$^b$                   & R$_{COLD}^c$   & R$_{ION}^c$ & F(Fe)$^d$ &  
$\chi^2_r$& N$_H^e$ & \\
\hline
TOTAL & 2.56$^{+0.07}_{-0.10}$ &  15.4$^{+0.3}_{-0.2}$ &  2.36$^{+0.03}_{-0.15}$  & 0.5$^{+0.3}_{-0.2}$ &0.7$^{+0.2}_{-0.2}$ & 1.57$^{+0.11}_{-0.37}$ &
   1.08 \\
INT 1 & 2.56$^{+0.05}_{-0.03}$ &  14.8$^{+0.4}_{-0.4}$ & 2.61$^{+0.21}_{-0.32}$  & 0.5$^{+0.3}_{-0.3}$ &0.4$^{+0.3}_{-0.3}$  & 1.59$^{+0.44}_{-0.28}$&    0.98 &
14.9$\pm0.2$\\
INT 2 & 2.57$^{+0.05}_{-0.04}$ &  18.8$^{+0.5}_{-0.3}$ &  2.38$^{+0.19}_{-0.05}$  & 0.4$^{+0.3}_{-0.3}$ &0.5$^{+0.4}_{-0.2}$  &1.51$^{+0.34}_{-0.28}$ &    0.95 &
19.4$^{+0.3}_{-0.2}$ \\
INT 3 & 2.57$^{+0.03}_{-0.03}$ &  14.6$^{+0.5}_{-0.5}$ &  2.77$^{+0.23}_{-0.54}$  & 0.5$^{+0.4}_{-0.4}$ &0.4$^{+0.4}_{-0.3}$  & $<1.3$&    1.03 &
14.7$\pm0.3$\\
\hline
\end{tabular}}
\footnotesize{$^a$: absorbing column density, 
in units of $10^{22}$~cm$^{-2}$. $^b$: Continuum normalization at 1~keV in units of 10$^{-2}$~photons~cm$^{-2}$~s$^{-1}$~keV$^{-1}$. $^c$: Ratio between the normalizations of the reflected and primary components. $^d$: 
Normalization of the narrow iron emission line, in units of 10$^{-2}$~photons~cm$^{-2}$~s$^{-1}$~keV$^{-1}$.
$^e$: $N_H$ values obtained by fitting the three intervals with a constant continuum, as discussed in Section~5.2.
}
\end{table*}

\begin{table*}
\caption{NGC 1365 - Time-resolved Spectral Fit: Iron Lines Parameters}
\centerline{\begin{tabular}{lccccccccccc}
\hline
OBS    & $E_N^a$            & $F_N^b$                    & $EW_N^c$           & $E_B^a$          & $F_B^b$              
       & $EW_B^c$           & $R_{IN}^d$                 & $R_{OUT}^d$        & $\Delta\chi^2$(S)$^e$ & $\Delta\chi^2$(G)$^e$  \\
\hline
TOT    &6.41$^{+0.04}_{-0.03}$ & 1.57$^{+0.11}_{-0.37}$     & 77$^{+5}_{-16}$ & 6.58$_{-0.06}^{+0.09}$ & 17.0$^{+1.3}_{-2.5}$ 
       & 535$^{+40}_{-80}$     & 4.5$^{+0.3}_{-0.2}$        & $>120$           & 30                 & 25                   \\
INT 1  &6.41$^{+0.04}_{-0.04}$ & 1.59$^{+0.44}_{-0.28}$     & 71$^{+28}_{-13}$ & 6.58$_{-0.08}^{+0.08}$ &  16.9$^{+2.1}_{-2.3}$ 
       & 485$^{+60}_{-65}$     & 4.5$^{+0.8}_{-1.2}$        & $>45$            & 30                 & 26                   \\
INT 2  &6.41$^{+0.03}_{-0.04}$ & 1.51$^{+0.34}_{-0.28}$     & 77$^{+17}_{-14}$ & 6.58$_{-0.07}^{+0.06}$ &  14.5$^{+4.0}_{-3.5}$                     
       & 326$^{+90}_{-80}$     &  $<3.4$                    & $>25$            & 15                  & 6                   \\
INT 3  &6.41$^{+0.08}_{-0.12}$ & $<1.3$                     & $<56$            & 6.58$_{-0.09}^{+0.14}$ &  16.9$^{+1.7}_{-2.2}$ 
       & 470$^{+50}_{-60}$     & $3.2^{+2.4}_{-0.5}$         & $>20$            & 18                  & 16                   \\
\hline
\hline
\end{tabular}}
\footnotesize{$^a$: Line peak energy, in keV. $^b$: Normalizations of the narrow (subscript $N$)
and broad (subscript $B$) emission lines,
in units of $10^{-5}$~keV~s$^{-1}$~cm$^{-2}$~keV$^{-1}$. $^c$: Equivalent width of the broad ($B$) and narrow ($N$) emission lines,
in eV.
$^d$: inner and outer radii of the broad line emitting region, in units of gravitational radii. $^e$: $\chi^2$ increase after
replacing the rotating black hole disk line with a non-rotating black hole disk component (S) and a Gaussian line (G). The Gaussian line peak
energy was left free, with best fit values in the range 5.8-6~keV.
}
\end{table*}

The only continuum parameter varying significantly is the 
absorbing column density $N_H$, which changes by $\sim3\times10^{22}$~cm$^{-2}$ ($\sim20$\%). 
The continuum slopes and normalizations are constant within the errors.
We plot in Fig.~5 the best fit values of these three parameters. 
The parameters related to
the absorption system at 6.7-8.3~keV, and the broad iron line feature
also remain constant within the errors, as well as the
flux of the narrow iron emission line. 
Moreover, the fluxes of the cold reflection components (continuum and narrow iron line) are
compatible with the values measured in all the previous X-ray observations of NGC~1365.


\begin{figure}
\epsscale{0.8}
\plotone{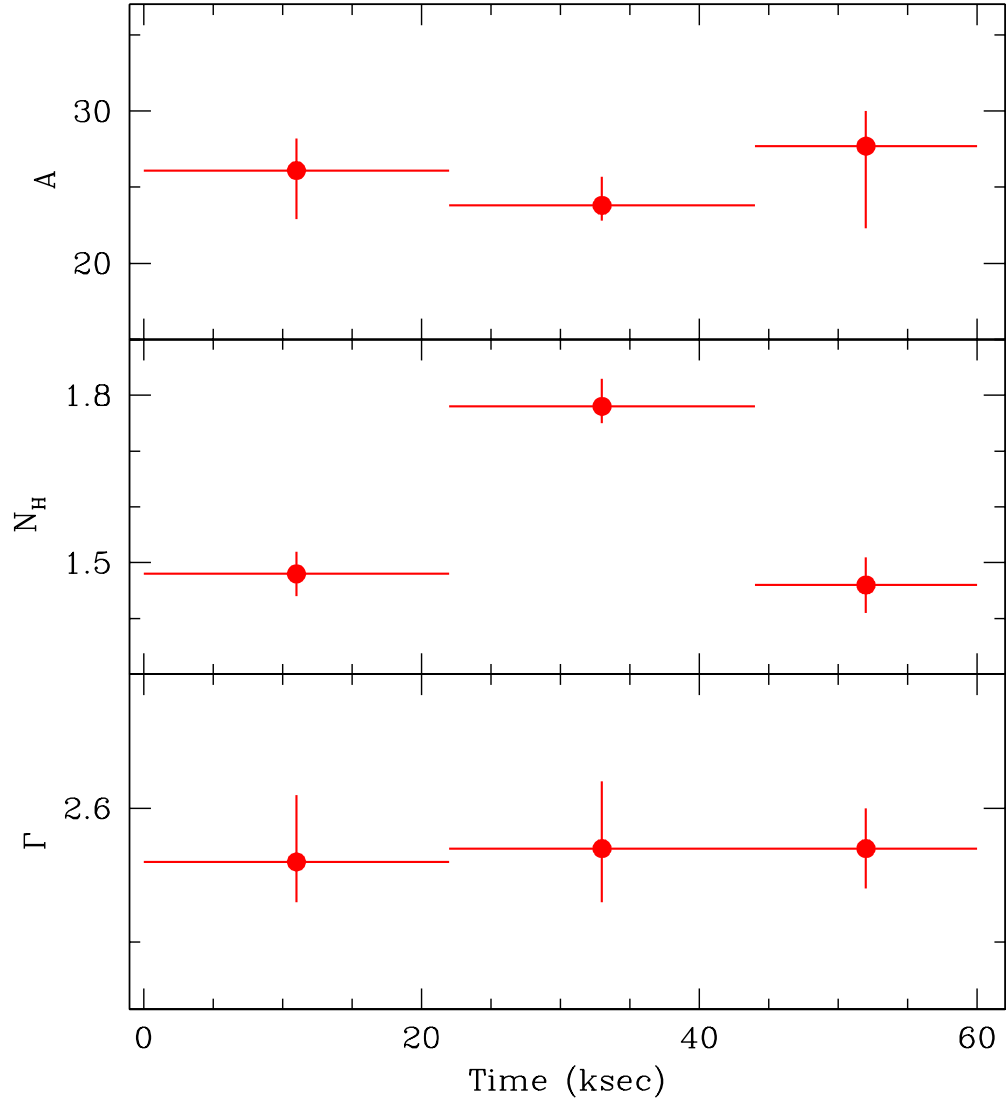} 
\figcaption{\footnotesize{
Results of the analysis of the spectra obtained from the three time intervals shown in Fig.~3. 
Upper panel: Continuum normalization at 1~keV,
in units of $10^{-3}$~keV~s$^{-1}$~cm$^{-2}$~keV$^{-1}$. Central panel: $N_H$, in units of $10^{23}$~cm$^{-2}$.
Lower panel: spectral slope $\Gamma$ of the primary continuum emission.
}}
\label{fit1}
\end{figure}

\subsection{Spectral fits with constant continuum}
The best fit models of the three intervals demonstrate that the observed  hardness ratio variations
are due to changes of column density along the line of sight. 
We therefore repeated the analysis fitting the three intervals simultaneously,
forcing all the parameters except the column density to have the same value in each interval.
The results of this new fit are completely compatible with those shown in Table~2, with
a factor $\sim$2 decrease of the uncertainties on the best fit values of the column densities:
$N_H(1)$=$14.9^{+0.2}_{-0.2}\times10^{22}$~cm$^{-2}$, 
$N_H(2)$=$19.4^{+0.2}_{-0.3}\times10^{22}$~cm$^{-2}$, 
$N_H(3)$=$14.7^{+0.3}_{-0.3}\times10^{22}$~cm$^{-2}$. 
Based on these results, we conclude that the statistical significances of the column density 
variations are at the level of $\sim20\sigma$  between the first and the second interval, 
and between the second and the third interval,
while the $N_H$ values are equal within the errors in the first and third interval.  

The best fit models obtained here have been used as starting points for the analysis 
of the two main subjects of this work: the column density variability on even shorter time scales, and the broad emission feature. We discuss these two points in the next two Subsections.

\subsection{Analysis of column density variations}

The obvious physical scenario reproducing the variability found in the spectral fits to the individual intervals 
is that of a cloud crossing the line of sight
to the X-ray source. If this is the case, a more physical model can be easily tested: instead of a 
single variable column density, the observed variations should be reproduced by a constant column density
covering the whole source, plus a second absorber with constant column density but a variable covering factor.

Moreover, 
the constancy of the continuum component on the timescales of the three intervals ($\sim20$~ks each)
suggests that the primary continuum remains constant during the whole observation at all timescales. It is 
highly unlikely that significant variations of the primary emission happened on time scales shorter than 20~ks,
since this would require that the average of these variations is almost exactly (within the measurement errors, which
are a few percent for the primary emission component, Table~2) the same in all the three intervals. Therefore, it is
safe to assume a constant primary emission at all time scales,
so that an analysis an even shorter time scale becomes possible.

We then performed a new spectral fitting, 
modifying two key aspects of our analysis: \\
1) We used a model with a constant continuum, absorbed by a constant
column density $N_H(TOT)$ plus a partial covering component, with a constant column density $N_H(PC)$ and a variable
covering factor $C(t)$. \\
2) 
We divided the 60~ks observation in 10 intervals of 6~ks each, and fitted them simultaneously  with
the above model. All the model components are required to have the same values in all the intervals, except for
the partial covering $C(t)$. 

The final results for the column density are $N_H(PC)$=$34.8^{+4.6}_{-1.4}\times10^{22}$~cm$^{-2}$ 
and $N_H(TOT)$=$12.3^{+0.1}_{-0.2}\times10^{22}$~cm$^{-2}$, 
while the results for the covering fraction $C$ are shown in Fig.~6. The variations of $C$
are highly significant and 
follow the expected pattern for an eclipsing cloud.
All the other parameters remain compatible within the errors with those shown in Table~2.

\begin{table}
\caption{Partial covering model}
\centerline{\begin{tabular}{lc}
\hline
\multicolumn{2}{c}{Continuum}\\
$\Gamma$                                            &  $2.49^{+0.01}_{-0.02}$ \\
$A^a$ (10$^{-2}$~keV cm$^{-2}$ s$^{-1}$ keV$^{-1}$)  &  2.75$^{+0.03}_{-0.04}$ \\
R$_{COLD}^b$                                               &  0.3$^{+0.2}_{-0.1}$ \\
R$_{ION}^b$						   &  0.4$^{+0.1}_{-0.1}$ \\
\hline
\multicolumn{2}{c}{Absorption} \\
$N_H$(TOT) ($10^{22}$~cm$^{-2}$)                     & 34.8$^{+4.6}_{-1.4}$  \\
$N_H$(PC)  ($10^{22}$~cm$^{-2}$)                     & 12.3$^{+0.1}_{-0.2}$  \\
$C_1$	     & 0.11$^{+0.02}_{-0.03}$  \\
$C_2$	     & 0.28$^{+0.02}_{-0.02}$  \\
$C_3$	     & 0.51$^{+0.01}_{-0.02}$  \\
$C_4$	     & 0.59$^{+0.02}_{-0.01}$  \\
$C_5$	     & 0.62$^{+0.01}_{-0.02}$   \\
$C_6$	     & 0.65$^{+0.01}_{-0.02}$   \\
$C_7$	     & 0.37$^{+0.02}_{-0.02}$   \\
$C_8$	     & 0.36$^{+0.02}_{-0.03}$   \\
$C_9$	     & 0.14$^{+0.02}_{-0.02}$   \\
$C_{10}$     & 0.47$^{+0.02}_{-0.02}$   \\
\hline
\multicolumn{2}{c}{Emission lines}\\
$E_N^c$ (keV)     & 6.41$^{+0.02}_{-0.02}$ \\     
$F_N^c$ (10$^{-5}$~keV cm$^{-2}$ s$^{-1}$ keV$^{-1}$)      & $1.8^{+0.2}_{-0.3}$ \\            
$EW_N^c$ (eV)      & 75$^{+10}_{-8}$ \\ 
$E_B^d$ (keV)     & 6.68$^{+0.03}_{-0.03}$ \\ 
$F_B^d$ (10$^{-4}$~keV cm$^{-2}$ s$^{-1}$ keV$^{-1}$)    & $ 2.1^{+0.2}_{-0.2}$ \\    
$EW_B^d$ (eV)      & 340$^{+35}_{-35}$ \\
$R_{IN}$ ($R_G$)    & 2.7$^{+0.2}_{-0.2}$ \\        
$R_{OUT}$ ($R_G$)   &  8.5$^{+3}_{-1}$\\
$\theta^e$ &  24$^{+8}_{-4}$\\
$q^f$      &  4.3$^{+0.5}_{-0.4}$\\
$\Delta\chi^2$(G)$^g$   &  25 \\
$\Delta\chi^2$(S)$^g$   &  70 \\
\hline
F(2-10)$^h$ (10$^{-11}$ erg cm$^{-2}$ s$^{-1}$)  & 1.24  \\
L(2-10)$^i$ (10$^{42}$ erg s$^{-1}$)            & 2.50   \\
$\chi^2$/d.o.f.  & 2687/2748  \\
\hline
\end{tabular}}
\footnotesize{$^a$: Normalization at 1~keV of the primary power law component.
$^b$: Ratio between the normalizations of the reflected and primary components.
$^c$: Narrow emission line.
$^d$: Broad emission line.
$^e$: Disk inclination angle in the broad line component.
$^f$: Slope of the disk emissivity profile.
$^g$: Increase of $\chi^2$ for a Gaussian model (G) or a Schwarzschild model (S) for
ther broad line. In the Gaussian model the line peak energy is free to vary, and its best  
fit value is $\sim5$~keV.
$^h$: Average observed 2-10~keV flux.
$^i$: Absorption-corrected 2-10~keV luminosity}
\label{tab_pc}
\end{table}

\begin{figure}
\label{covering}
\plotone{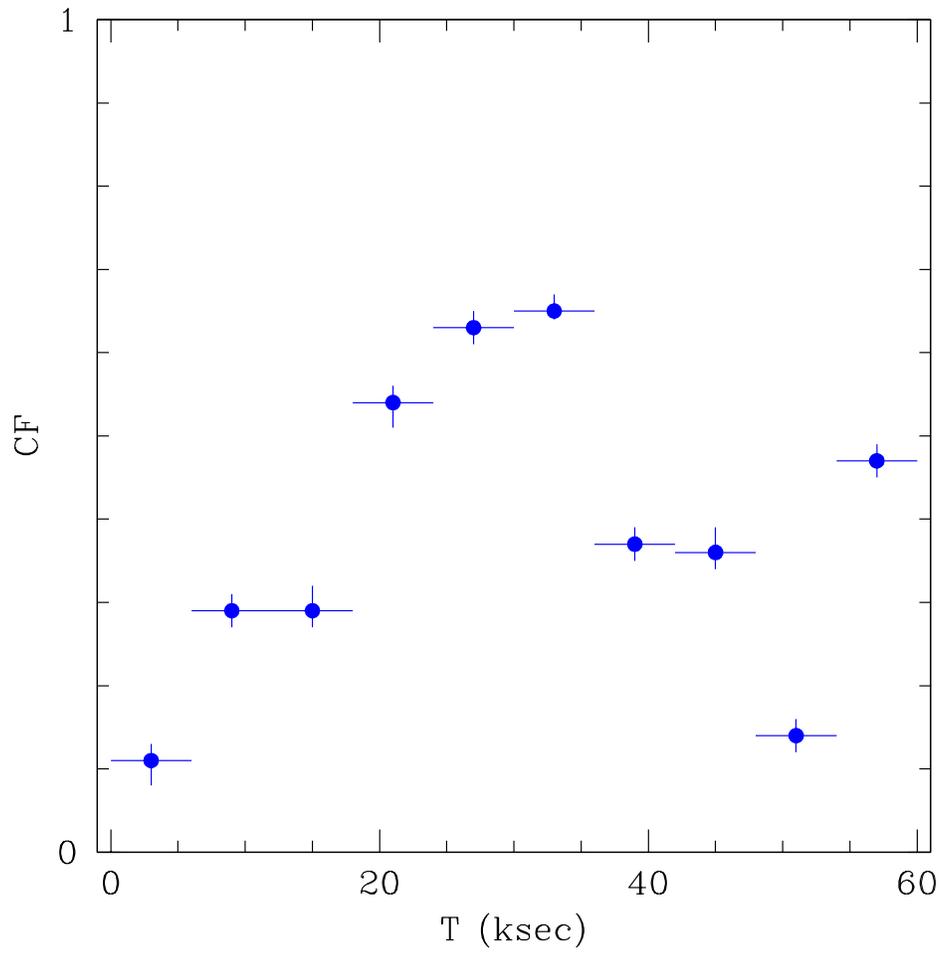} 
\figcaption{\footnotesize{
Light curve of the covering fraction of the partial covering component, obtained by splitting the total
exposure in 10 6-ks intervals.
}}
\end{figure}

We note that the amount of information contained in Fig.~6 is much higher than in Fig.~5: 
our time-resolved spectral analysis resolves a complete transit of a cloud in front of the X-ray source.
The precise measurement of the crossing time allows us to put stringent limits on the dimensions of the X-ray source and
the obscuring cloud, and on the distance of the cloud from the central black hole. We will discuss this issue in detail in 
Section~6.

Since this model, obtained through the steps discussed above, is our final "best interpretation" of
the time and spectral behavior of NGC~1365, we report all the relevant fit parameters
in Table~4 and we show the main components in Fig.~7.

The relatively low errors on the single values of the partial covering parameter are obviously dependent on
our choice of adopting a constant continuum. Since the analysis of the three long intervals presented in Section~5.1
shows that the continuum remains constant with a maximum statistical uncertainty of $\sim10$\%, 
we repeated our analysis allowing continuum variations in each
interval up to 10\%. We found fully consistent values for all the parameters, with the absolute errors on the
covering factors higher by $\sim30$\%. This does not affect any of our conclusions.

For completeness, we mention that a similarly good fit is obtained replacing the partial covering absorber
with a single, fully covering absorber with variable column density. The best fit values obtained
in this scenario are all consistent with those shown in Table~4. The $N_H$ best fit values range from 14 to 20$\times10^{23}$~cm$^{-2}$,
with relative errors of a few percent in each interval. 
We cannot distinguish between these two scenarios on statistical grounds. However, the partial covering
model is the most likely scenario from a physical point of view. 

\begin{figure}
\label{covering}
\plotone{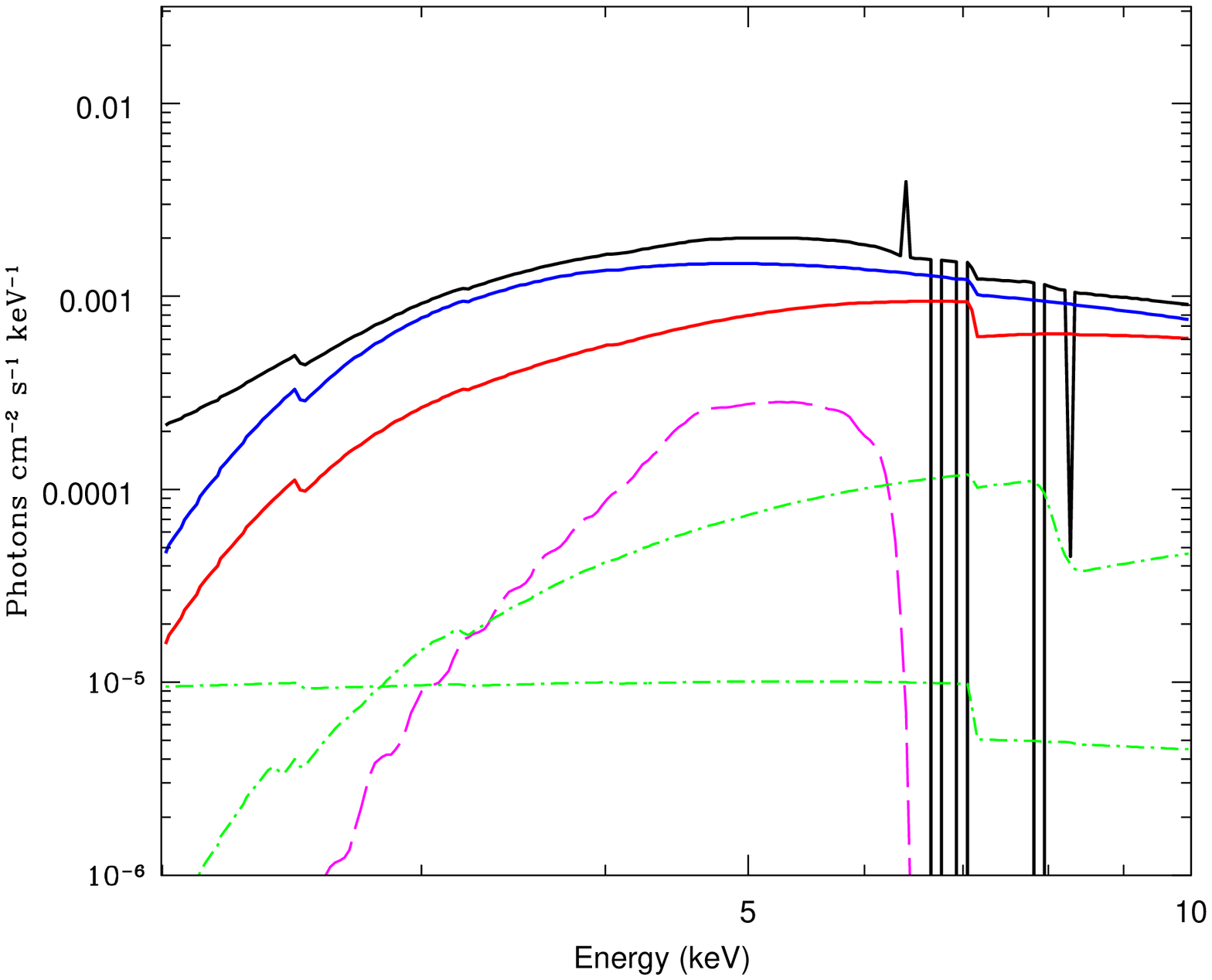} 
\figcaption{\footnotesize{Main components of our final, variable partial covering model. Upper continuous line (black):
total model, including the narrow absorption and emission lines. Bottom continuous lines: absorbed primary
power law, for the minimum (blue) and the maximum (red) covering factor in Table~4. Dashed line (magenta): relativistic
iron emission line. Dot-dashed lines (green): ionized and neutral reflection components. 
}}
\end{figure}

\subsection{Analysis of the broad iron feature}

A broad emission feature at energies between 4 and 7 keV is clearly present in the spectra (Fig.~2) and
strongly requested in all our fits (Table~3 and~4). Even allowing for a variable partial covering, and a ionized reflector, as
in the model described in the previous Section, the requirement for this feature remains strong: the fit improvement
with respect to a narrow component only in our final fit is $\Delta\chi^2=205$ with four additional parameters. 

We also note that a simple broad Gaussian line model provides a significantly worse fit: 
$\Delta\chi^2=20$, with a best fit peak energy
$E\sim5.7$~keV. If the peak energy is forced to be within the physically acceptable interval 
(6.39-6.9~keV) the fit is worse by
$\Delta\chi^2=160$. Moreover a model assuming a relativistic profile from a non-rotating 
black hole also provides a worse
fit ($\Delta\chi^2=70$). 

In order to visualize these results, we plot in Fig.~8 the relevant data regarding the broad emission line.
A plot of each interval is not easily readable, because the statistics is too low, and the information
is hidden in the noise. Therefore, we plot the total residuals to the best fit without the relativistic line 
component, obtained by adding the residuals of each individual spectrum. For comparison, we also plot the 
residuals to the best fit {\em with} the line component, and the model components representing the
intrinsic (unabsorbed) line and the observed (absorbed) line for the first (least obscured) and fifth (most obscured)
interval. These model lines contain interesting information:  they show the non-negligible effects of absorption
on the broad line, and show the difference between the different phases of the eclipse. 

\begin{figure}
\epsscale{0.8}
\plotone{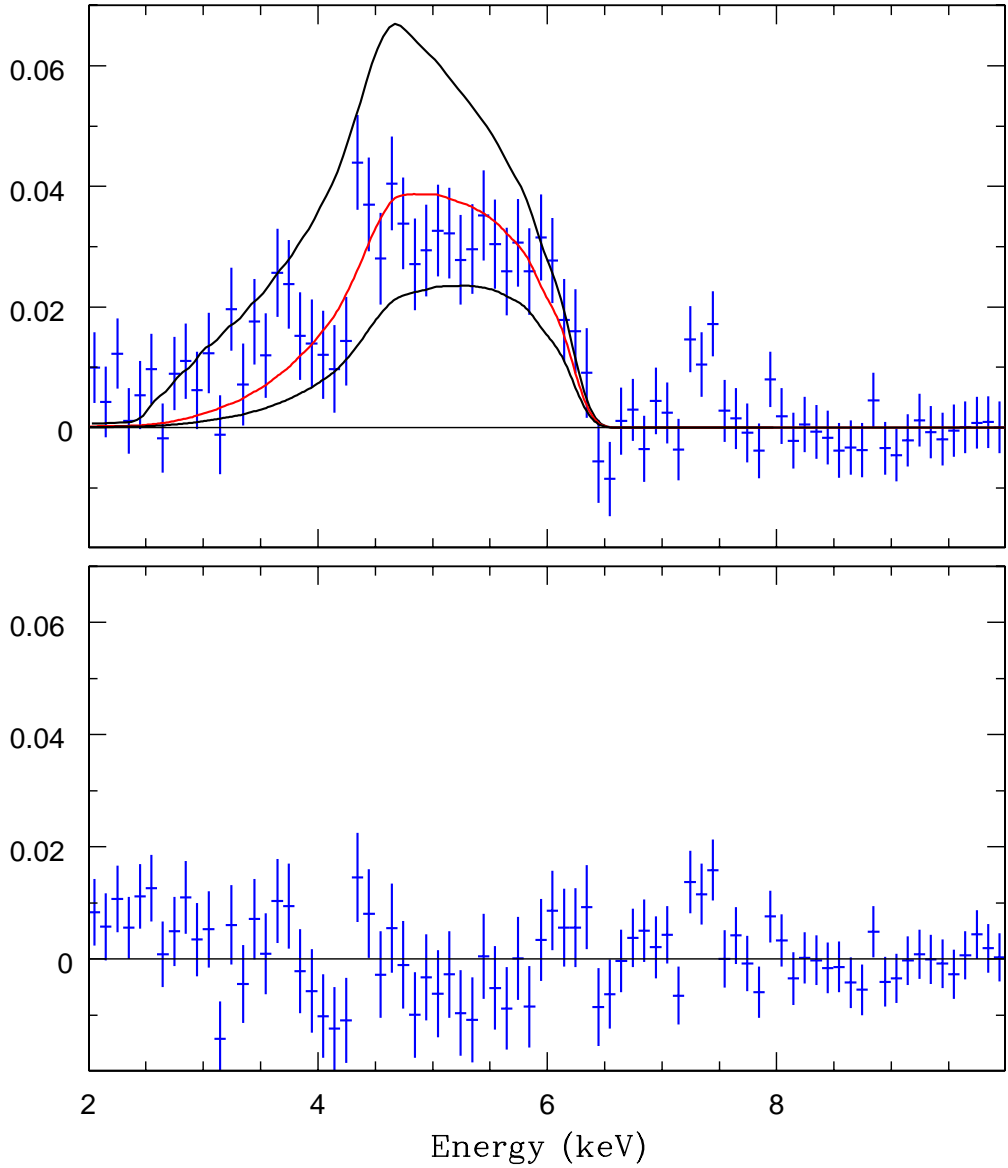} 
\figcaption{\footnotesize{
{\em Upper panel}: ratio to the best fit model without the broad iron line components. 
The profile has been obtained
by adding the residuals of all the spectra obtained from the 10 intervals in which we split the observation.
The black, highest line shows the best fit unabsorbed line. The two lower lines show the observed (i.e. absorbed)
line profiles for the first and the fifth interval, where the covering factor of the absorbing cloud was
minimum and maximum, respectively. {\em Lower panel}: as above, residuals to the complete best fit model, including
the relativistic iron line. }}
\label{iron2}
\end{figure}

As a further step in order to obtain a physically consistent scenario, we repeated our fits using a relativistic blurring model
applied to the whole disk reflection spectrum, i.e. to both the disk emission line and the warm 
reflection continuum component, assumed to originate from the disk as well. 
The relativistic blurring has been applied to the
disk reflection model of Ross \& Fabian (2005) which includes both the reflection continuum and the iron emission line
from a ionized disk. 
This replaces the blurred ionized reflection and the LAOR broad emission line components in our model. All the other components are unchanged with respect to the 
 previous Subsection.

The main results are the following:\\
$\bullet$ The overall fit is similar to the standard partial-covering one shown in Table~4 ($\chi^2$/d.o.f.=2700/2748)\\
$\bullet$ The best fit values for the inner and outer radii of the emitting region are 
$R_{IN}=5^{+1}_{-1} R_G$ and $R_{OUT}=14^{+8}_{-4} R_G$. \\
$\bullet$ The best fit ionization parameter is $U=1.1^{+0.1}_{-0.3}\times10^3$ erg$\times$cm/s.\\
$\bullet$ The iron abundance is $2.8^{+0.2}_{-0.3}$ with respect to standard solar abundances. 
\\
$\bullet$ The best fit values of the other parameters are compatible with those found in the simpler partial-covering fit,
shown in Table~4.

We note that in the previous model, while the line emitting region parameters (emissivity profile, inner and outer radius,
inclination angle) are well determined
by the fit, the analogous parameters for the continuum components are not significantly constrained, as already noted
in Section~2.1. Since we now apply the relativistic blurring to a line+continuum model, it is natural to 
obtain parameters in agreement with those of the LAOR model in the previous fit. The main advantage in this
new model is a consistent treatment of the relative strength of the line and continuum components, which allows
an estimate of the iron abundance, not possible in the ``standard'' partial covering model of Table~3.

\subsection{Results from other XMM-Newton observations}

Two more {\em XMM-Newton} observations of NGC~1365 have been performed since 2004: a further 60~ks
observation performed $\sim$6~months after the one presented here, in July~2004, and a ``long look''
consisting of three consecutive orbits ($\sim370$~ks total) in 2007. Here we report the
results of the analysis of both. We just quickly summarize the results on the ``long look'', which are
presented in another paper (Risaliti et al.~2008), while we give some more details on the analysis of the
second 60~ks observations, which is not presented elsewhere.

{\bf 1. The long look}. An {\em XMM-Newton} 3-orbit monitoring of NGC~1365 was performed in May~2007.
During this long observation the source was mostly in a Compton-thick, reflection-dominated state. 
However, during the second orbit we observed a ``hole'' in the Compton-thick absorber, which revealed
the intrinsic X-ray source. 
A detailed analysis of the emission and absorption components, as the one presented here, is not possible for
this observation, due to the high residual column density in the Compton-thin state 
($N_H$$\sim9$$\times$20$^{23}$~cm$^{-2}$). Therefore, a much simpler model has been adopted,
without
the broad line and absorption lines components.
However, the duration of the ``anti-eclipse'' allowed a precise determination
of the size of the X-ray source and of the distance of the Compton-thick clouds. These results, presented
in Risaliti et al.~2008, are complementary to our current analysis of the Compton-thin absorber, in order
to obtain a complete view of the circumnuclear absorber/reflector of NGC~1365.
 
{\bf 2. The July~2004 60~ks observation}. A further 60~ks {\em XMM-Newton} observation (hereafter OBS~2) was performed in
July~2004, six months after the observation discussed in this paper (hereafter OBS~1).
We applied to these data the same analysis described in the previous Sections, starting from a fit of the
whole observation and then searching for possible spectral variations. The main relevant results of this
analysis are the following:\\
- The Light curves of both the total emission and the hardness ratio do not show significant variations.
As a consequence, we only analyzed the spectrum obtained from the whole observation.\\
- The source is in a Compton-thin state with an equivalent column density $N_H\sim3\times10^{23}$~cm$^{-2}$
(see below for more details on the structure of the absorber) .
This implies that the red tail of the broad iron emission line is heavily suppressed by absorption.
We included a relativistic broad line in our model, and we found that this component is
perfectly consistent with that found in the OBS~1 spectrum. However, the  uncertainties on the
line parameters are too high to add additional evidence to the case of the broad relativistic line.
Therefore, we will not discuss this component further.

The continuum emission can be fitted equally well with two different models: (1) a fully obscured
power law with $\Gamma\sim2.2$ and $N_H\sim3\times10^{23}$~cm$^{-2}$ (this is the model 
presented in R05B), and (2) an obscured power law with $\Gamma\sim2.5$ and $N_H\sim2.5\times10^{23}$~cm$^{-2}$
partially obscured by a second absorber, with $N_H\sim2.5\times10^{23}$~cm$^{-2}$, and covering fraction
$C=70\pm10$\%. The two fits are statistically equivalent ($\chi^2_{red}=1.01$ for both). This means that the
inclusion of the second, partially covering absorber does not improve the fit from a purely statistical
point of view. However, the best fit continuum parameters make the partial covering model particularly interesting:
not only the photon index is the same as in OBS~1, but the power law normalization is also perfectly
consistent ($A=2.5^{+0.3}_{-0.4}\times10^{-2}$~erg~cm$^{-2}$~s$^{-1}$~keV$^{-1}$). All the other model components
(reflection continuum, emission and absorption lines) are also consistent with the OBS~1 parameters shown in Table~4.
Since all the fitting parameters, and in particular the slope and normalization of the primary continuum,
are left free to vary, these results suggest a scenario with a constant continuum on time scales
of months, and with the absorber as the only variable component. Obviously, we cannot exclude that
the simpler, one-absorber model is the correct one, and that the constancy of the continuum parameters 
assuming the partial covering model is due to a ``cosmic conspiracy''.
Future studies, involving the analysis of all the available X-ray observations of NGC~1365 through 
partial-covering models, will clarify this interesting issue.
However, in the present paper we consider the partial covering model as the most likely from a physical point of view.

\section{Discussion: The variable absorber}
The obscured AGN in NGC~1365 has several unique properties, which make it the best laboratory to investigate
the structure of the obscuring/reflecting medium (both disk and circumnuclear gas) found so far among obscured AGNs.
In the previous Section we presented a detailed time-resolved spectral analysis of a 60~ksec {\em XMM-Newton} 
observation, which showed a clear variation of obscuration due to a cloud 
with $N_H\sim3\times10^{23}$~cm$^{-2}$ crossing the line of sight.

The partial eclipse of the X-ray source allows a direct estimate of the sizes
of the X-ray emitting region and of the obscuring cloud, and of the distance between them.
From the analysis of Fig.~5 and Table~4 we note that the time profile of the variation 
cannnot have a flat top longer than about 12~ks, being the values of $C_5$ and $C_6$ consistent, 
while $C_5$$>$$C_4$ and $C_6$$>$$C_7$. This suggests that
the covering cloud is about of the same size as the X-ray source. If the cloud were
much bigger than the source, it should cover a roughly constant fraction of the source in the central phase of the eclipse,
leading to a plateau in the covering factor light curve. On the other hand, the cloud section cannot
be smaller than $\sim65$\% of the source size to achieve the maximum covering factor ($C_6\sim65$\% in Table~4). 
A more detailed study of the shape of the obscuring cloud and
of the emitting region is not possible with the available data. 
A higher S/N detection of an eclipse could in principle provide strong constraints
on the geometry.
This could be possible with
future X-ray observatories, or, even using currently available instrumentation, if an eclipse 
with a higher $N_H$ cloud and/or with a complete obscuration of the central source is observed.
This would greatly increase the
precision of the measurements of the covering factors, thus allowing a more detailed investigation of the shapes
of cloud and source.
     
In order to estimate the size of the X-ray emitting region,  
we assume a spherical, homogeneous cloud with diameter $D_C$, column density
$N_H$ and number density
  $n=N_H/D_C$, at a distance $R$ from the central black hole, moving with Keplerian velocity $V_K=\sqrt{GM_{BH}/R}$.
Since the observed eclipse is almost complete, we assume that the linear size of the X-ray source is $D_S\sim V_K\times T$,
where $T$ is the measured occultation time, which in our case is $\sim$25~ks. An estimate of $V_K$ can
be obtained from three different arguments:

1) Limits on the ionization parameter of the obscuring cloud. The observed absorption variations require that the
obscuring cloud is not overionized.
We require $U<U_{MAX}=$20~erg$\times$cm/s (see Risaliti et al.~2007, 2008 for details). 
Using this limit, and the geometrical condition that the cloud must be of the same size of the X-ray source
(see discussion above),
after straightforward algebra we
obtain $R>(GM_{BH})^{1/5}[T\times L_X/(U_{max}N_H)]^{2/5}$. The black hole mass can be estimated using the
mass-bulge luminosity, mass-bulge velocity dispersion, and mass-H$\beta$ width relationships.  
The available estimates are
 log$(M_{BH}/M_\odot)=7.3\pm0.4(0.3)$ from the $M_{BH}$-bulge velocity dispersion
correlation (Oliva et al.~1995, Ferrarese et al.~2006) and log$(M_{BH}/M_\odot)=7.8\pm0.4(0.3)$ 
from the relation between $M_{BH}$ and the
K magnitude of the host bulge (Dong \& De~Robertis~2006, Marconi \& Hunt~2003)
 where the errors include statistical and systematic
effects, and the number in brackets refer to the statistical dispersion of the correlation.
We note that the $M_{BH}$-bulge luminosity correlation likely overestimates $M_{BH}$ in the case of
NGC~1365, due to the high starburst contribution to the bulge luminosity.
A much smaller estimate (but still compatible with the $M_{BH}$-velocity dispersion relation) is obtained from
the width of H$\beta$: assuming FWHM(H$\beta$)=1,900~km/s, as measured by Schulz et al.~1999, and the Kaspi et al.~(2005)
$M_{BH}$-FWHM(H$\beta$) relation, we obtain $M_{BH}$$\sim$2$\times$10$^6$~$M_\odot$.

Adopting 
$L_X$=1.9$\times10^{42}$~erg~s$^{-1}$~cm$^{-2}$
(obtained by extrapolating our absorption-corrected luminosity to the 1-100~keV band), and the measured
$N_H$ and $T$, we obtain R$>$2$\times10^{15} M_7^{1/5}$~cm, $V_K$<$7,600 M_7^{3/10}$~km~s$^{-1}$, 
and $D_S$$<$$1.9\times10^{13} M_7^{3/10}$~cm, where $M_7$=$M_{BH}/(10^7 M_\odot)$. 
This size corresponds to 6 gravitational radii for $M_{BH}$=$10^{7}~M_\odot$, and 20 gravitational radii for 
$M_{BH}$=2$\times$10$^6 M_\odot$.

2) Geometrical limits. A purely geometrical limit comes from the requirement that the cloud and the X-ray source
are physically separated, i.e. that the distances from their centers is $R>R_{min}=(D_C+D_S)/2\sim D_S$. From $D_S=V_K\times T$
we obtain $R_{min}$=$(GM_{BH})^{1/3} T^{2/3}$=9.4$\times10^{13} M_7^{1/3}$~cm, $D_S<R_{min}$ and
$V_K<38,000 M_7^{1/3}$~km~s$^{-1}$. These values describe an unlikely geometry, with the edge of the cloud touching
the edge of the X-ray source. However, they are interesting because they represent strong limits to the 
source size. We note that even in this extreme configuration, the size of the emitting region is only $\sim$30 gravitational radii for $M_{BH}$=10$^7 M_\odot$.

3) Limits from the width of the emission line. We can reasonably assume that the variable component of the 
obscuring gas is the same responsible for
the cold reflection component present in our best fit models, and best seen in the previous observations of NGC~1365 which caught the
source in a Compton-thick, reflection dominated state. We can therefore use the width of the ``narrow'' iron emission line 
to estimate the velocity of the obscuring clouds, under the assumption that the thick clouds (responsible for the bulk of
the iron emission line) and the thin ones are at about the same distance from the central source.
We note that a small contribution to the observed narrow emission line
can be given by the constant $N_H$ component, for which no constraints on the distance based on our arguments are possible.
However, since the column density of this component is only of the order of $10^{23}$~cm$^{-2}$, we expect 
only a minor contribution to the observed iron line flux. The narrow iron line is resolved in the ``long look'' observation
presented in Risaliti et al.~2008, which gives
$V_{Fe}=2^{+1}_{-1}\times10^3$~km~s$^{-1}$. Considering that the
measured width of the emission line is affected by the projection effects of a distribution of clouds rotating around the central source,
while the crossing cloud is moving in the plane of the sky, we obtain $V_K=3000$~km~s$^{-1}$, $R\sim1.5\pm0.7\times10^{16}$~cm, and
 $D_S<7.5\times10^{12}$~cm=2.5~$M_7^{-1} R_G$. This low value suggests that the correct value for the black hole
mass is not higher than log($M_{BH}$/$M_\odot$)=6.5, in agreement with the $M_{BH}$-FWHM(H$\beta$) correlation. 

These results are complemented by the analysis of the 60~ks July~2004 observation, summarized in Section~5.5,
and the 370~ks ``long look'' (Risaliti et al.~2008). 

The July~2004 observation shows that the Compton-thin absorber is not only made of clouds of the same size
of the X-ray source, as observed in the January~2004 observation. 
In particular, the constancy of the covering fraction $C$ in the July~2004 observation implies that the covering cloud
is significantly larger than the X-ray source. Assuming a cloud with radius $D_C$, covering at most
$\sim60$\% of the source, in order to have C constant within 15\%, as observed (Section~5.5), we must have $D_C>3 D_S$.
An alternative way to obtain the same constancy is by assuming a distance of the obscuring cloud from the center 
$\sim10$~times larger than in the January~2004 observation. We consider this possibility unlikely, though still possible.

The ``long look'' provides complementary information of the Compton-thick phase of the absorber, showing that
clouds with $N_H>10^{24}$~cm$^{-2}$ are also present, at about the same distance from the central source as the
thin clouds. The persistence of the Compton-thick state of $>$2~days also suggests that the thick clouds 
are significantly larger than the X-ray source (even if a superposition of several, smaller thick clouds
cannot be excluded, see Risaliti et al.~2008 for details).

Summarizing, the available {\em XMM-Newton} observations of NGC~1365 suggest that the obscuring medium is
made of clouds with column densities from a few~$10^{23}$~cm$^{-2}$ to $>10^{24}$~cm$^{-2}$, and sizes
form about that of the X-ray source to several times larger.

\section{Discussion: The broad iron line}

The broad iron line found in our observation is best fitted by a relativistically smeared profile, due to a  rotating
black hole. The significance of the broad line is one of the highest ever found among AGNs. 

The fit with a rotating black hole is preferred with respect to the pure Schwarzschild case. However, the
relatively large inner radius obtained in our fits ($R_{IN}\sim3$ and $R_{IN}\sim4-6$ in the 
two final fits discussed above) suggests that a maximally rotating black hole disk is not really requested by our data.
More detailed models of the line profiles (e.g. Dovciak et al.~2004) show that solutions with a a maximally-rotating Kerr profile
and an inner radius $R_{IN}\sim4$ are equivalent (with our data quality) to solutions starting from the
least stable orbit and with an angular momentum parameter $a\sim0.6$. Since we cannot resolve
this degeneracy, we conservatively conclude that our data suggest that the black hole is rotating with $a>0.6$.


The detailed fits performed in order
to study the variability of the circumnuclear absorber are a strong check on the significance of this feature.
Indeed, our models contain several components which could mimic a relativistic line: a cold and a 
ionized reflection component, and a variable partial covering. In particular, the variable partial 
covering introduces a curvature in the integrated
continuum which could be interpreted as due to the red tail of the relativistic line. 
It is quite instructive to compare the line parameters obtained from the total 60~ks observation (Table~3) 
with those obtained from the time-resolved fit (Table~4):\\
- The (absorption corrected) equivalent width 
of the broad line in the former is significantly higher than in our final best fit including partial covering. 
Its value is higher than the typical equivalent width measured in Seyfert 1s by a factor of at least 3.
Such high values are suspicious, and are the indication of a possible 
incorrect estimate of the continuum level in the global fit, due to not properly considering the
spectral curvature due to the $N_H$ variations. \\
- The parameter errors are significantly larger. In particular, the outer radius of the emitting line is not well constrained,
and the uncertainty on the line flux is of the order of 20\%, a factor $\sim3$ higher than in our final fit.\\
All the problems present in the fit of the whole observation are solved with the time-resolved approach. 
Our final best fit still requires a relativistic iron line, and the correct estimate of the variable continuum level
provides a more reasonable equivalent width (Table~4) and a better determination of the line parameters.


Overall, the variable partial-covering models (the standard one with separate iron and reflection components, and the
Ross \& Fabian~(2005)  one where the line and continuum reflection
are treated self-consistently), provide a complete view of the X-ray emitting region and the reflecting/absorbing medium:

- The equivalent width of the line is higher than 300~eV, among the highest values measured for these features in 
nearby AGNs (e.g. Miller et al.~2007). This high value is physically explained by the high iron metallicity estimated in
the final line+continuum reflection model (Z(Fe)$\sim3$). 

- The outer radius of the line emitting region, estimated both from the relativistic line component in the standard partial
covering model and from the relativistic blurring component in the {\em kdblur} model, is relatively small 
($R_{OUT}<15~R_G$). This parameter is interesting when compared with the size of the X-ray source:
both the eclipses observed during the {\em Chandra} campaign and the {\em XMM-Newton} long look, 
and the occultation event occurred during this observation suggest that the
X-ray source is confined within a few gravitational radii from the central black hole. 
The upper limit on the outer radius of the iron line is a further, completely independent  indication supporting this 
scenario. 

- In addition to the broad iron line and the extreme variability behavior, the AGN in NGC~1365 shows another peculiar property,
not analyzed here (but included in all the models): 
the system of four iron absorption lines in the 6.7-8.3~keV interval, discussed in detail in R05B.
The main results of the analysis of those lines are that they are due to a highly ionized ($U>1000$erg$\times$cm/s) absorber, with a 
column density of a few $10^{23}$~cm$^{-2}$. The physical properties of this gas strongly suggest that it is located
very close to the central source (at $\sim50-100~R_G$), i.e. ``inside'' the clumpy cold absorber.
These properties strongly suggest that this gas is the Compton-thin, lower density 
extension at larger radii of the warm, Compton-thick component responsible for the iron emission line and the
highly ionized reflection. In this scenario, the expected emission of this component would be
a weak scattering component, not distinguishable from the direct continuum emission in the spectral fit.


- The high equivalent width of the four absorption lines, never observed in other similar AGNs, suggests some peculiar property
of this absorber, not present with the same intensity in other sources. The most obvious reason for the presence of these
strong lines, beside a particularly high column density and high ionization warm absorber (see above), is an 
overabundance of iron with respect to the other heavy elements. This would easily explain the unusual strength of the observed lines:
a factor of $\sim$a few enhancement of iron in NGC~1365 would justify the lack of similar detections in other bright AGNs.
This scenario is now strongly supported by our measurement of the equivalent with of the broad emission line, a factor
$\sim$3 higher than the average value measured in local AGNs, and by the best fit iron abundance in the relativistic
blurring-partial covering model, Z(Fe)$\sim3$. At completely different scales, an iron overabundance is also suggested
by the analysis of the {\em XMM-Newton}-RGS high resolution soft X-ray spectrum of the diffuse emission (Guainazzi \& Bianchi~2007,
Guainazzi et al.~2008, in prep.). In order to further check the case for an iron overabundance,
we also fitted the cold absorber with a variable iron abundance.
However, we were not able to obtain strong constraints for this extra parameter, which can be anyway consistent with
an iron overabundance of $\sim3$ without any significant change in the fits presented in the previous Sections. 
Summarizing, the available observations strongly suggest the presence
of an iron overabundance at all scales from $\sim1$~kpc down to a few gravitational radii form the central black hole.

Despite the consistency of the above scenario, the interpretation of the broad 4-6~keV feature as a
relativistic iron line is not unique. In particular, it has been proven that the same profile can be reproduced
by a combination of partially ionized variable absorbers (Miller et al.~2008). Here we do not fully analyze this scenario, 
but we only note two important aspects: 1) in principle,  
it is not possible to rule out an extra continuum plus warm absorber(s) component
with the same profile as our emission line, if multiple components with free ionization states are allowed; 
2) none of our absorbing/reflecting components seem to introduce alterations in the continuum shape that have not been
included in our fits. In particular, we used the XSTAR code in XSPEC to simulate
 the possible contribution of the warm absorber associated to the 
Fe~XXV and Fe~XXVI absorption lines. 
The result is that a gas with an ionization parameter as high as the one measured from the lines' peak
energies and flux ratios (R05B) does not significantly affect the continuum profile. In order
to produce significant spectral curvature, a lower ionization state is required, which would produce a detectable set
of absorption structures around 6.5-6.6~keV, none of which have even been detected in NGC~1365.
  
\section{Conclusions}

We have presented a complete, time resolved spectral analysis of a 60~ks {\em XMM-Newton} observation of 
the AGN in NGC~1365.
The source was on average in a high flux, relatively low-absorption ($N_H\sim1.5\times10^{23}$~cm$^{-2}$) state.
An analysis of the hardness ratio light curve revealed a strong spectral variability during the observation,
which we interpret as due to a cloud with $N_H=3.5\times10^{23}$~cm$^{-2}$ crossing the line of sight in $\sim25$~ks,
with a maximum coverage of the X-ray emitting region of about 65\%.

The analysis also revealed the presence of a strong, broad iron emission line,
best fitted with a 
relativistic profile due to a rotating black hole. The significance of this line has been confirmed allowing
for neutral and ionized reflection, and for time variable partial absorption. The best fit outer radius 
of this line is of the order of $\sim10~R_G$. This is an  indication that the X-ray emitting region is concentrated 
in the inner part of the accretion disk, in agreement with the results of the variability analysis: the parameters of the 
observed eclipse suggest that the X-ray source is confined within $<20~R_G$ from the central black hole.

The main conclusions of our study are the following:
\begin{itemize}
\item The partial eclipse detected in this observation demonstrates that the obscuring material consists
(at least in part) of clouds with column density of the order of a few $10^{23}$~cm$^{-2}$. Assuming that these
clouds rotate with Keplerian velocity around the central source, we obtain a distance from the central black hole 
of the order of 
10$^{16}$~cm. The inferred cloud density is a few 10$^{10}$~cm$^{-3}$. These are typical values for
clouds emitting the high ionization broad emission lines in the optical/UV of type~1 AGNs. 
Therefore, we identify the variable circumnuclear absorber as due to a Broad Line Region cloud.
\item The inferred size of the X-ray emitting region is of the order of $10^{13}$~cm. Assuming the black hole
mass value obtained from the standard relations between black hole mass and bulge luminosity/velocity dispersion,
this corresponds to $\sim10-25$~gravitational radii. 
\item We fitted a global model for the accretion disk reflection 
(both line and continuum component) smeared by the general relativistic
effects of a rotating black hole, and found a iron overabundance of a factor of $\sim3$. This high value
easily explains (a) the high significance of our detection, "helped" by the unusually high equivalent width of
the broad line, and (b) the peculiar set of iron~XXV and iron~XXVI absorption lines present in this spectrum
(and discussed in a previous paper, R05B), with an optical depth never measured in any other (bright) AGN.
With a standard solar iron abundance both features would be barely detectable.
\item By comparing the results of our last, time-resolved spectral analysis with the standard global spectral fit, we notice
that the parameters relative to these components are significantly different in the two fits, and the uncertainties are much
larger in the global fit. In particular, the best fit  line equivalent width is higher in the latter ($EW\sim500$~eV),
and the inner and outer radii are poorly constrained. This can be explained if the absorption variations introduce a 
curvature in the continuum global spectrum, which is incorrectly fitted with an excessively strong and broad emission line.
As a consequence, the case of variable partial absorption should be always be checked in the spectral analysis of
long X-ray observations of AGNs.
\end{itemize}

The results presented here, together with those from the {\em Chandra} monitoring campaign (Risaliti et al.~2007),
and the {\em XMM-Newton} long look (Risaliti et al.~2008) provide a detailed view of the circumnuclear absorber,
 both in the Compton-thin and Compton-thick components.

However, more work is still to be done on the X-ray emission of NGC~1365: indeed,
 in order to constrain the structure and size of the X-ray emitting region and of the circumnuclear clouds,
the ``optimal'' observation is still missing: 
a continuous monitoring of a complete eclipse of a low-$N_H$, Compton-thin state, 
by a Compton-thick cloud. Such observation would better constrain the properties of the
circumnuclear absorber and would offer a unique opportunity to perform 
an experiment of ``accretion disk tomography'': at different phases of the eclipse,
different regions of the X-ray source would be visible.

Finally, a further on-going development of our study of the circumnuclear
medium of AGNs consists of applying the time-resolved analysis presented here
to more sources. Our first choices are the few other known ``changing look''
AGNs, i.e. sources which were observed in both a Compton-thin and a reflection-dominated
state. For one of these objects, UGC~4203 (the ``Phoenix Galaxy'', Guainazzi et al.~2002),
we already obtained a {\em Chandra} monitoring campaign analogous to that performed
on NGC~1365. The (positive) results of these observation will be presented in a future paper.

\acknowledgements

We are grateful to the referee for helpful comments which significantly improved the
clarity of the paper.
This work was partially supported by NASA grants 
G06-7102X and NNX07AR90G.



\begin{thebibliography}{}
\bibitem[Arnaud(1996)]{1996ASPC..101...17A} Arnaud, K.~A.\ 1996, 
Astronomical Data Analysis Software and Systems V, 101, 17 
\bibitem[Bianchi et al.(2005)]{2005MNRAS.357..599B} Bianchi, S., Matt, G., 
Nicastro, F., Porquet, D., \& Dubau, J.\ 2005, \mnras, 357, 599
\bibitem[Bianchi et 
al.(2006)]{2006A&A...448..499B} Bianchi, S., Guainazzi, M., \& Chiaberge, M.\ 2006, \aap, 448, 499 
\bibitem[Dong \& De Robertis(2006)]{2006AJ....131.1236D} Dong, X.~Y., \& De Robertis, M.~M.\ 2006, \aj, 131, 1236
\bibitem[Elvis et al.(2004)]{2004ApJ...615L..25E} Elvis, M., Risaliti, G., 
Nicastro, F., Miller, J.~M., Fiore, F., 
\& Puccetti, S.\ 2004, \apjl, 615, L25 
\bibitem[Ferrarese et al.(2006)]{2006ApJ...644L..21F} Ferrarese, L., et 
al.\ 2006, \apjl, 644, L21 
\bibitem[Ghisellini, Haardt, \& Matt(1994)]{1994MNRAS.267..743G} 
Ghisellini, G., Haardt, F., \& Matt, G.\ 1994, \mnras, 267, 743 
\bibitem[Gilli et al.(2000)]{2000A&A...355..485G} Gilli, R., Maiolino, R., 
Marconi, A., Risaliti, G., Dadina, M., Weaver, K.~A., \& Colbert, E.~J.~M.\ 
2000, \aap, 355, 485
\bibitem[Guainazzi(2002)]{2002MNRAS.329L..13G} Guainazzi, M.\ 2002, \mnras, 
329, L13 
\bibitem[Guainazzi 
\& Bianchi(2007)]{2007MNRAS.374.1290G} Guainazzi, M., \& Bianchi, S.\ 2007, \mnras, 374, 1290 
\bibitem[Iyomoto et al.(1997)]{1997PASJ...49..425I} Iyomoto, N., Makishima, 
K., Fukazawa, Y., Tashiro, M., \& Ishisaki, Y.\ 1997, \pasj, 49, 425 
\bibitem[Kaspi et al.(2005)]{2005ApJ...629...61K} Kaspi, S., Maoz, D., 
Netzer, H., Peterson, B.~M., Vestergaard, M., 
\& Jannuzi, B.~T.\ 2005, \apj, 629, 61
\bibitem[Krolik \& Begelman(1988)]{1988ApJ...329..702K} Krolik, J.~H.~\&
Begelman, M.~C.\ 1988, \apj, 329, 702
\bibitem[Laor(1991)]{1991ApJ...376...90L} Laor, A.\ 1991, \apj, 376, 90 
\bibitem[Magdziarz \& Zdziarski(1995)]{1995MNRAS.273..837M} Magdziarz, 
P.~\& Zdziarski, A.~A.\ 1995, \mnras, 273, 837 
\bibitem[Marconi \& Hunt(2003)]{2003ApJ...589L..21M} Marconi, A.~\& Hunt, 
L.~K.\ 2003, \apjl, 589, L21 
\bibitem[Matt, Guainazzi, \& Maiolino(2003)]{2003MNRAS.342..422M} Matt, G., 
Guainazzi, M., \& Maiolino, R.\ 2003, \mnras, 342, 422
\bibitem[Miller et 
al.(2008)]{2008A&A...483..437M} Miller, L., Turner, T.~J., \& Reeves, J.~N.\ 2008, \aap, 483, 437 
\bibitem[Miller(2007)]{2007ARA&A..45..441M} Miller, J.~M.\ 2007, \araa, 45, 441 
\bibitem[Oliva et al.(1995)]{1995A&A...301...55O} Oliva, E., Origlia, L., 
Kotilainen, J.~K., \& Moorwood, A.~F.~M.\ 1995, \aap, 301, 55
\bibitem[Peterson \& Wandel(2000)]{2000ApJ...540L..13P} Peterson, B.~M.~\& 
Wandel, A.\ 2000, \apjl, 540, L13 
\bibitem[Puccetti et al.(2007)]{2007MNRAS.377..607P} Puccetti, S., Fiore, 
F., Risaliti, G., Capalbi, M., Elvis, M., 
\& Nicastro, F.\ 2007, \mnras, 377, 607 
\bibitem[Risaliti, Maiolino, \& Bassani(2000)]{2000A&A...356...33R} 
Risaliti, G., Maiolino, R., \& Bassani, L.\ 2000, \aap, 356, 33 
\bibitem[Risaliti(2002)]{2002A&A...386..379R} Risaliti, G.\ 2002, \aap, 
386, 379 
\bibitem[Risaliti, Elvis, \& Nicastro(2002)]{2002ApJ...571..234R} Risaliti, 
G., Elvis, M., \& Nicastro, F.\ 2002, \apj, 571, 234 
\bibitem[Risaliti et al.(2005)]{2005ApJ...623L..93R} Risaliti, G., Elvis, 
M., Fabbiano, G., Baldi, A., \& Zezas, A.\ 2005a, \apjl, 623, L93 (R05A)
\bibitem[Risaliti et al.(2005)]{2005ApJ...630L.129R} Risaliti, G., Bianchi, 
S., Matt, G., Baldi, A., Elvis, M., Fabbiano, G., 
\& Zezas, A.\ 2005b, \apjl, 630, L129 (R05B) 
\bibitem[Risaliti et al.(2007)]{2007ApJ...659L.111R} Risaliti, G., Elvis, 
M., Fabbiano, G., Baldi, A., Zezas, A., 
\& Salvati, M.\ 2007, \apjl, 659, L111 
\bibitem[Risaliti et al.(2008)]{2008arXiv0811.1594R} Risaliti, G., et al.\ 
2008, arXiv:0811.1594
\bibitem[Schulz et 
al.(1999)]{1999A&A...346..764S} Schulz, H., Komossa, S., Schmitz, C., M{\"u}cke, A.\ 1999, \aap, 346, 764
\bibitem[Str{\"u}der et 
al.(2001)]{2001A&A...365L..18S} Str{\"u}der, L., et al.\ 2001, \aap, 365, L18 
\bibitem[Turner et 
al.(2001)]{2001A&A...365L..27T} Turner, M.~J.~L., et al.\ 2001, \aap, 365, L27 
\bibitem[Urry \& Padovani(1995)]{1995PASP..107..803U} Urry, C.~M.~\&
Padovani, P.\ 1995, \pasp, 107, 803
\bibitem[Wang et al.(2009)]{2009arXiv0901.0297W} Wang, J., Fabbiano, G., 
Elvis, M., Risaliti, G., Mazzarella, J.~M., Howell, J.~H., 
\& Lord, S.\ 2009, arXiv:0901.0297 

\end{thebibliography}
\end{document}